\documentclass{article}
\usepackage{amsfonts}
\usepackage{amssymb}
\usepackage{amsmath}
\usepackage{amsthm}
\usepackage[english]{babel}
\begin{document}

\title{\bf Modified Gravitational Collapse, \\ or the Wonders of the MOND}

\author{{\bf Alexey Golovnev} \quad and \quad {\bf Natalia Masalaeva}\\
{\small {\it Saint-Petersburg State University, high energy physics department.}}\\
{\small \it Ulyanovskaya ul., d. 1; 198504 Saint-Petersburg, Petrodvoretz; Russia.}\\
{\small golovnev@hep.phys.spbu.ru \qquad nataxa.masalaeva@mail.ru}}
\date{}

\maketitle

\begin{abstract}

There are many hot discussions in the literature about two competing paradigms in galactic and extra-galactic astronomy and cosmology, namely the Dark Matter and the Modified Newtonian Dynamics (MOND). It is very difficult to challenge MOND from the cosmological side because a full relativistic realisation is needed in the first place, and any failure can then be attributed to a particular model, and not to the MOND itself. We propose to study non-relativistic stages of gravitational collapse in MOND which, we argue, is a relevant task for this competition. Spherically symmetric dust cloud collapse and intrinsic unavoidable non-linearities of the deep MOND regime are discussed. We conclude that complicated, both numerical and {\it analytic}, studies of modified gravitational dynamics are needed in order to assess the viability of MOND.

\end{abstract}

\vspace{1cm}

\section{Introduction}

The paradigm of MOND -- Modified Newtonian Dynamics -- was created \cite{MOND} in order to account for the notorious plateau in the peripheral parts of the galactic rotation curves. 

Those curves have clearly shown that, given the standard laws of mechanics and gravity, there must reside far more masses in galaxies than it could have been inferred from the luminous matter alone. The most straightforward solution is to assume that there is some, yet unknown, dark matter which makes the rotation curves as flat as they are. If we further assume that the Dark Matter is cold\footnote{or may be warm but definitely not hot} and very weakly interacts with ordinary "baryonic" matter, then it perfectly fits into the gross picture of the cosmological evolution. It is a breath-taking achievement because this scientific chef-d'{\oe}uvre is tightly constrained by the expansion history data, primordial nucleosynthesis yields, observations of the CMB radiation, dynamics of the large scale structure formation, and the baryon acoustic oscillations. On the other hand, the physical nature of the CDM -- Cold Dark Matter -- particles is not known yet; and the non-observation of supersymmetric partners at the LHC cannot pour any new enthusiasm into the issue.

A very different approach is known under the name of MOND. On its first appearance, it claimed that the luminous mass is practically the only mass that is there, and therefore the laws of mechanics and/or gravitation must become modified. As it was, it could only concern the framework of rotation velocities in galaxies, the very setup which it has been invented for. And it is the field in which the paradigm has witnessed its largest success, most notably in reproducing the (baryonic) Tully-Fisher relation. But clearly, a model can be truly predictive only when it comes out of its own cradle. It was absolutely necessary to find a proper physical realisation for the ideas of MOND. In the non-relativistic limit, the mostly accepted one goes in terms of the non-linear Poisson equation \cite{BekensteinMilgrom}. For a full relativistic theory, there are many possibilities with the most advanced ones coined in the tenets of {\it TeVeS} \cite{TeVeS}. The expectations from full relativistic models are rather high. Indeed, as grandiose as it may sound, they are expected to reproduce the whole success of modern cosmology with all its observational data, and even a bit more.

So, it is indeed very remarkable that the MONDian paradigm could ever come anywhere close to this highly ambitious aim \cite{MONDcosmology}. However, the overall performance still falls short of being satisfactory, and it is even less obvious that, at the end of the day, a more elegant and simpler solution would be proposed as compared to the standard Dark Matter model \cite{Glenn}. A common attitude to the failures of {\it TeVeS} is that they only talk against some particular realisations, and not the MOND paradigm in general. In a sense, this is certainly true, but does not give much credibility to the whole endeavour if we are to {\it understand} the Universe.

It is very important to find some bridges which may connect the cosmological challenges for MOND with its basic undoubted assumptions. We believe that the problem of gravitational collapse might serve such a role. We show that the gravitational collapse in MOND has some peculiar features, and it may provide an additional input of information together with the studies of structure formation in {\it TeVeS} \cite{MONDcosmology} and numerical N-body simulations in MOND \cite{Nusser, Llinares, Angus}. Balancing at the bridge can in principle show the extent to which the problems of MONDian cosmology might be traced back to the basic principles of low acceleration dynamics. In Section 2 we briefly review the two competing paradigms, Dark Matter and MOND. In Section 3 we discuss the gravitational collapse of a (spherically symmetric) dust cloud. In Section 4 we dwell a bit upon the intrinsic non-linearity of MOND, although it has been discussed many times since the very birth of the subject. And, finally, in Section 5 we conclude.

\section{DM vs MOND -- Much Ado}

The puzzle of the hidden mass has been discovered about 80 years ago by applying the virial theorem to the Coma Cluster. And the X-ray luminosity of galaxy clusters shows presence of very hot intergalactic gas which, being bound inside the cluster, implies a much deeper gravitational well than the one that can be provided by gravitational attraction of the visible matter alone. Inside the galaxies, rotation curves asymptote to a very flat plateau in their peripheral parts instead of Newtonian $v\sim \frac{1}{\sqrt{r}}$ behavior. The standard conclusion which stemmed out of that hurdle was the existence of huge amounts of invisible matter in galaxies and galaxy clusters. And by now, assuming the laws of General Relativity, we know that, though a substantial amount of baryonic matter is hidden from our eyes, no more than 5 percent of the total energy budget of the Universe lies within our current understanding of physics while one quarter is given by unknown weakly interacting heavy Dark Matter particles with remaining 70 percent in even more mysterious form of Dark Energy. We can measure these fractions via the cosmic history, dynamics of expansion, the spectrum of temperature fluctuations in the CMB, and the baryon acoustic oscillations scale inferred from large galaxy surveys. The resulting picture agrees nicely with constraints from primordial nucleosynthesis and with the need of cold matter for successful structure formation.

In MOND, a new fundamental constant of Nature is introduced, the fundamental acceleration $$a_0\approx 1.2\cdot 10^{-8}\ \rm\frac{cm}{s^2}$$ such that for $a\gg a_0$ the dynamics is Newtonian while for $a\ll a_0$ the gravitational force scales with distance as $\frac{1}{r}$ instead of Newtonian $\frac{1}{r^2}$. Clearly, the $\frac{1}{r}$ dependence of the centripetal acceleration explains the constant linear velocity plateau in galactic rotation curves. As a non-relativistic law, this idea cannot be directly applied to cosmology. However, for galaxies it makes a good job. And surprisingly, the unique acceleration constant $a_0$ appears to describe the whole range of known galaxies quite well. This is just how it must be in MOND but does not look natural for the Dark Matter cosmology. Moreover, since 1977 we know the so-called Tully-Fisher relation \cite{TF} for the luminosities of spiral galaxies, $L\sim v_{\infty}^4$, where $ v_{\infty}$ is the rotational velocity at the plateau. It is not that straightforward to explain this relation in the standard paradigm. But what is even more amusing, substitution of the {\it baryonic} mass\footnote{to be honest, {\it visible} baryonic mass (stars and hot interstellar gas)} instead of the luminosity makes it even better satisfied by the astronomical data \cite{bTF}. This relation can be deduced almost trivially in MOND. Indeed, the gravitational force law for a test mass $m$ at distance $r$ from a gravitating body of mass $M$ reads
\begin{equation}
\label{MOND}
F=m\frac{\sqrt{GMa_0}}{r}
\end{equation}
in the deep MOND regime. The centripetal acceleration is given by $\frac{v_{\infty}^2}{r}$, and therefore $v_{\infty}=\sqrt[4]{GMa_0}$. As tautological as it may seem, it is very remarkable, even if only for its very tautology.

The competition of the two paradigms has sparkled many hot debates in the literature. One of the reasons is that there are indeed some serious problems within the standard cosmology when the properties of nearby galaxies are put under scrutiny \cite{Kroupa, Peebles}. To name just a few, they include the Local Group structure (the Local Void is too empty, and there are too many large galaxies on its outskirts), missing satellites problem (around the Milky Way and other nearby big galaxies), existence of pure disk (bulgeless) galaxies, the cuspy cores of predicted Dark Matter profiles, reported very high peculiar velocities \cite{DarkFlow} of galaxy clusters (Dark Flow), etc.  Of course, they generically refer to highly non-linear and complicated processes of formation of galaxies and galaxy clusters. At the same time, in general the $\Lambda\rm CDM$ cosmology has passed lots of independent tests with a clear and indisputable success. It is very tempting to see if the problems can be surmounted without giving up the Dark Matter. It is not  inconceivable that some problems are due to incomplete understanding of baryonic (ordinary matter) physics during the wild events of forming the galaxies. Some tensions might be ameliorated by warming up the Dark Matter fluid \cite{Menci}. And, for example, the Dark Flow may simply not be there \cite{noDarkFlow, Planck-noDarkFlow}... 

Some of the problems (Local Void, bulgeless disks) are arguably immune to adjusting the details of interactions of stars and gas \cite{Peebles}. They seem to call for making the attractive forces stronger during the structure formation \cite{Peebles}. But does it necessarily amount to modifying gravity, along the lines of MOND or in some other ways? One very attractive possibility is to introduce new interactions into the Dark Sector \cite{NGP}. We do not know the nature of Dark Matter, and there is absolutely no reason to assume that it exerts no non-gravitational self-interactions. The additional forces might also lead to higher peculiar velocities, modified density profiles, etc.

And, last but not least among the problems, the impressively good performance of MOND in what concerns the observations of galaxies is by itself a rather big puzzle for the Dark Matter model. What is the origin of the universal acceleration $a_0$ and all the regularities and simple phenomenological laws of galactic dynamics? Why the baryonic Tully-Fisher relation works so good \cite{bTF}? Why the central halo surface density was reported \cite{Paolo} so constant and universal for all conceivable types of galaxies? Sure, if it is to be the order born from chaos, then it would be anything but unique. However, it does not overthrow the necessity to explain the outcome.

The most important task for MOND is to extend its own success beyond the galactic rotation curves. This is non-trivial because it was just a phenomenological construction. We have to specify the dynamical laws which lead to the force (\ref{MOND}). In non-relativistic limit the community has largely agreed upon the modified Poisson equation\footnote{with $|\nabla\phi|\to 0$ at $r\to\infty$ as the boundary condition}
\begin{equation}
\label{modPoisson}
\nabla\left(\mu\left(\frac{|\nabla\phi|}{a_0}\right)\cdot\nabla\phi\right)=4\pi G \rho,
\end{equation}
where $\mu(x)$ is a smooth function which is meant to interpolate between the Newtonian $\left.\mu(x)\right|_{x\gg 1}\to 1$ and deep MOND $\left.\mu(x)\right|_{x\ll 1}\to x$ limits. It is disappointing to have an arbitrary function here, but actually the simplest choice of $\mu(x)=\frac{x}{\sqrt{1+x^2}}$ works well. Note that this is a highly non-linear equation, and a naive expectation to find the MONDian acceleration in terms of the Newtonian one $a_N$ by simple prescription
\begin{equation}
\label{simpleMOND}
a_N=\mu\left(\frac{a}{a_0}\right)\cdot a
\end{equation}
is incorrect because we can add a curl term without influencing the equality in (\ref{modPoisson}), and generically we {\it must} add it for making the acceleration a gradient. A simple version of MOND which modifies the second law precisely according to (\ref{simpleMOND}) is possible but comes at a high price: momentum conservation is violated (consider two interacting particles in regions with different $x\equiv\frac{a}{a_0}$). However, in spherical symmetry, the Stokes' theorem shows that the two models coincide. There is also the so-called QUMOND (quasi-linear MOND) \cite{QUMOND} proposal which posits that the MONDian potential is given by equation of the form 
\begin{equation}
\label{QUMOND}
\bigtriangleup\phi=\nabla\left(\nu\left(\frac{|\nabla\phi^{(N)}|}{a_0}\right)\cdot\nabla\phi^{(N)}\right)
\end{equation}
with Newtonian potential $\phi^{(N)}$ and a smooth function $\nu(x)$. It requires a more involved action principle with both $\phi$ and $\phi^{(N)}$ inside, but nicely combines the tractability of simple MOND (\ref{simpleMOND}) with the less pathological character of canonical MOND (\ref{modPoisson}). The N-body simulations \cite{Llinares} indicate that the difference between canonical MOND and simple MOND is considerable but not overwhelming.

The first failure comes very soon. It is true that the gravitational force (\ref{MOND}) rather slowly depends on the distance, but it is also true that its increase with the attracting mass is equally slow. In its regime of validity, it is always stronger than the Newtonian force but, for a ball of uniform density, it grows as $\sqrt{r}$ instead of the linear growth in Newtonian gravity, and therefore becomes closer and closer to the latter case until the Newtonian regime sets back again. For the galaxy clusters, the MONDian force is not strong enough. And so, some amount of Dark Matter arises in MOND \cite{clusters}. Of course, it can all be baryonic, and in the standard cosmology we need even much more invisible baryonic matter than that. Some even argue that it might be due to (relatively large) neutrino masses. But it is still disappointing for the paradigm which was created in order to avoid the need of introducing chaotic dark components for explaining the simple and regular laws. Given this fact, one would probably expect that the model should have problems with creating enough large scale structures. Surprisingly, the N-body simulations have, on the contrary, exhibited problematically {\it increased} galaxy clustering in simple MOND \cite{Nusser, Llinares}, canonical MOND \cite{Llinares} and QUMOND \cite{Angus}. We will come to this point back later.

One more strike is a clear visual confirmation of Dark Matter in the Bullet Cluster \cite{Clowe}. The gravitational lensing signal and the X-luminous hot gas are spatially segregated in the process of clusters' collision, consistently with Newtonian gravity. Of course, they also have some amount of Dark Matter in MOND, and it is probably not impossible to simulate the same situation in pure MOND \cite{BulletMOND}. But is it worth doing so if the heavy dark matter seems being a simpler and more elegant solution? Well, currently it would be unfair to definitely tell against MOND on the basis of the Bullet Cluster because its kinematics presents a severe problem for $\Lambda\rm CDM$. The collisional velocities are too high \cite{badBullet}. Probably, an interacting dark sector could make a better job than both paradigms do.

Direct experimental tests of MOND seem absolutely unfeasible in our large gravitational acceleration environments unless we assume the strong equivalence principle and a free falling reference frame. In a recent paper \cite{Das} it was shown that in a free falling laboratory with strong equivalence principle it might be possible to test the simple MOND law (\ref{simpleMOND}) experimentally. The choice of the most pathological version of MOND is probably not that important. However, the equivalence principle is absolutely crucial for the argument. And it is very unlikely that it holds in any viable realisation of MOND \cite{MOND, BekensteinMilgrom}. Therefore, cosmology remains the only arena for testing the paradigm.

Of course, to compete with the standard cosmology, the model must have some saying on cosmological problems. For that we need a fully relativistic realisation of MOND. There is much less agreement on this issue in the MONDian community. However, it seems reasonable to take up the most advanced proposal, the {\it TeVeS} \cite{TeVeS}. The first, purely scalar-tensor, version \cite{BekensteinMilgrom} of it has failed badly because the scalar field cannot enforce additional light bending to reproduce the correct gravitational lensing by galaxies. These services got assigned to the vector sector. In summary, the scalar field exerts the required fifth force onto the non-relativistic matter particles with the two limits of $\mu(x)$ provided by a special non-canonical kinetic term of the scalar, and the vector interactions are tuned to ensure the correct bending of light.

In this realisation, the linear cosmological perturbation theory has been developed and confronted with the CMB temperature spectrum and the spectrum of baryon density perturbations \cite{MONDcosmology}. For a reasonable cosmology to emerge, the cosmological constant and the massive neutrino component with $\Omega_{\nu}\approx 0.15$ were introduced. But nevertheless, there remains a problem of the third peak hight in the CMB. At the same time, a satisfactory spectrum of linear baryon density fluctuations has been obtained. This is a very strong and important result. But it can be shown analytically \cite{Dodelson} that the vector degrees of freedom were instrumental in producing enough gravitational instability in {\it TeVeS}. Therefore, we come back to employing some dark new degrees of freedom practically unrelated to the baryonic content of the Universe. And there are still many more challenges from cosmology to be addressed.

\subsection{What is MOND?}

How should we frame our attitude towards MOND and its potential implications for our understanding of cosmology? The initial proposal was nothing more than a phenomenological description of the observed regularities in how the galaxies look like. No doubt, it was impressively successful on galactic scales, and this very fact calls for explanation whichever paradigm we stick to. And there are some intriguing and potentially far-reaching coincidences. For example, it was noted by Milgrom that $a_0$ has the same order of magnitude as $cH_0$ where $H_0$ is the Hubble constant, and $c$ is the velocity of light.

However, if MOND is to become a new paradigm for a fundamental theory of Nature, then we definitely need something more. We must provide a clear embedding into a coherent theoretical framework which includes non-stationary motions, strong gravitational fields, cosmological models, detailed interactions with matter. Many attempts were made focusing on these goals. As we have described above, there are certain successes, and MOND does enter a new era featured by very interesting proposals towards a full-fledged cosmological theory of MONDian type.

There are many unresolved problems and obstacles on this way, which is of course understandable. But regretfully, each failure gets ascribed to any particular realisation, and not to the MOND itself thus making the latter practically unbeatable outside its initial field of explaining the dynamical equilibrium in galaxies. There is some truth behind such reassignment, but in order to decide which of the two models is better we need to consider their overall performance in all the relevant aspects. And we know that the Dark Matter plays many roles, and its failures refer to the most complicated and non-linear part of the story, the structure and formation of galaxies and galaxy clusters, for which some yet unknown, or unaccounted, physics of both dark and baryonic sectors may become of paramount importance. 

Sometimes people even plot a "theory confidence graph" for the standard $\Lambda\rm CDM$ cosmology \cite{Kroupa} counting each problem as if it was a clear failure, independent of all others, and leading to a certain decline in confidence such that the latter gets claimed to be practically evanescent. Of course, this approach ignores lots of confirmed cosmological predictions. And if we use Dark Matter for some other problems of cosmology it cannot simply disappear when it comes to the realm of galaxies. 

On the other hand, MOND also requires some amount of invisible matter for description of galaxy clusters, and some new degrees of freedom when constructing relativistic realisations and building cosmologies. And it actually hits at the very heart of its statement. Why the dynamics in galaxies is so regular with the unique fundamental acceleration $a_0$? Why these rules fail for the clusters? If we need an extra mass in clusters, of the same order of magnitude as the visible one, why don't we see it in galaxies? If we were able to perform a reliable and detailed simulation of galaxy formation in MOND, would it do better about the problems which are claimed to rule out the Dark Matter? If not, then which model is more flexible, or are there any other options?

Our opinion is that, in totality, the phenomenological appeal of MOND and its fundamental theory embeddings is not better than that of the standard cosmological model. The theoretical foundations seem rather ad hoc. For example, in $\it TeVeS$ the two limits of $\mu(x)$ are provided by postulating the corresponding two limits of the non-linear kinetic function of the scalar field, and the correct bending of light is tuned by the vectors. At the same time, the Dark Sector might be successfully modified without destroying the rest of cosmology, and there are some works in that direction. For sure, we also need to explore all other possibilities including modifications of gravity. And in this respect, MOND is very useful in showing us what a modified gravity model must look like if it is to substitute the Dark Matter, or at least to account for a large part of its effect.

\section{Gravitational collapse}

In absence of any clear agreement on the correct fundamental theory behind the basic tenets of MOND, we would like to learn as many lessons directly from the latter as we can. The problem of gravitational collapse in a static flat background might be one of such lessons. Of course, in reality the structure formation proceeds in an expanding Universe, and it does make an important difference. However, for advanced stages of gravitational collapse the background dynamics can be thought of as subdominant, and one might also wonder to what extent the potential problems of structure formation can be traced back to the basic assumptions irrespective of any choice of relativistic extensions.

As the simplest possible model, we consider a spherically symmetric pressureless dust cloud. And as a simple estimate of the structure formation time, one can calculate the time of free fall from an initial radius $r=R$ to the centre. In Newtonian gravity we take the equation of motion $$\frac{d^2 r}{dt^2}=-\frac{GM(r)}{r^2}$$ under the initial conditions (Cauchy data) $r(0)=R$ and ${\dot r}(0)=0$ where $M(r)$ is the gravitating mass inside the ball of radius $r$ around the centre. We assume that the different layers of the cloud do not cross each other, and therefore the mass inside a given co-moving layer is always equal to its initial value $M$. Multiplying this equation of motion by $\dot r$, we get the conservation law ${\dot r}^2=2GM\left(\frac1r - \frac1R\right)$. Let us also assume that the dust cloud has a constant density $\rho$ from the centre to the border, and $M(R)=\frac43 \pi\rho R^3$. After that, the free fall time is obtained by a simple integration, 
\begin{equation}
\label{Newtonian time}
T_N=\sqrt{\frac{3}{8 \pi G\rho R^2}}\int\limits_{0}^{R}\frac{dr}{\sqrt{\frac{R}{r} - 1}}=\sqrt{\frac{3}{2 \pi G\rho}}\int\limits_{0}^{1}dy\sqrt{1-y^2}=\sqrt{\frac{3\pi}{32 G\rho}}.
\end{equation}
The result does not depend on the initial radius $R$. The Newtonian dynamical law just marginally ensures that the layers of different $r$ (all initially at rest) in a uniform density ball do not cross each other before reaching the central point.

Let us now see what happens in deep MOND. The equation of motion reads
\begin{equation}
\label{MONDfall}
\frac{d^2 r}{dt^2}=-\frac{\sqrt{GM(r)a_0}}{r},
\end{equation}
and the conservation law takes the form ${\dot r}^2=-2\sqrt{GMa_0}\cdot\ln\frac{r}{R}$. Again, assuming that the layers do not cross, we use $M(R)=\frac43 \pi\rho R^3$ and get
\begin{equation}
\label{MONDian time}
T_M=\sqrt[4]{\frac{3}{16\pi G\rho R^3 a_0}}\int\limits_{0}^{R}\frac{dr}{\sqrt{\ln \frac{R}{r}}}=\sqrt[4]{\frac{3R}{16\pi G\rho a_0}}\int\limits_{0}^{1}\frac{dy}{\sqrt{\ln \frac1y}}=\sqrt[4]{\frac{3R}{\pi G\rho a_0}}\int\limits_{o}^{\infty}dz\cdot e^{-z^2}=\sqrt[4]{\frac{3\pi R}{16 G\rho a_0}}.
\end{equation}
Now the free fall time goes as $R^{1/4}$, and the outer layers come to the centre with a delay. In central parts the collapse occurs faster than in the periphery. Note though that this formula is valid only as long as $a<a_0$. Equation (\ref{MONDfall}) gives the $a_0$ acceleration at $R_{\star}=\frac{3a_0}{4\pi\rho G}$. At the same time, $T_M$ becomes larger than $T_N$ at $R=\frac{3\pi a_0}{64 G\rho}=\frac{\pi^2}{16}R_{\star}$  which is slightly smaller than the radius $R_{\star}$ of transition to Newtonian regime. We calculate the ratio of the two free fall times, (\ref{Newtonian time}) and (\ref{MONDian time}),
\begin{equation}
\label{time ratio}
\frac{T_M}{T_N}=\sqrt[4]{\frac{64\rho GR}{3\pi a_0}}=\sqrt[4]{\frac{16}{\pi^2}\cdot\frac{a_N}{a_0}}
\end{equation}
where $a_N<a_0$ is the Newtonian acceleration at the initial time at $r=R$, and see that we can make the MONDian free fall time a bit larger than the Newtonian one. Of course, even this marginal increase shows that, in reality, the free falling MONDian layers eventually enter Newtonian regime because the MONDian laws postulate that the force can only be larger than in purely Newtonian dynamics which cannot lead to enlargement of the free fall time. A good message is that the collapse can occur considerably faster than in Newtonian gravity for a sparse cloud with $a_N\ll a_0$. This is probably the reason for successful structure formation in MOND despite the absence of Dark Matter \cite{Nusser, Llinares, Angus}.

The dynamics of collapse is different from that in Newtonian gravity where the whole uniform cloud shrinks to zero size for all layers independently of the radius. In MOND, the inner layers come to the centre first, previously entering the Newtonian regime.  One can also see that from density profiles. Assuming that layers do not cross, we have $\frac{d}{dt}\left(4\pi r^2(t)\rho(t)\delta r(t)\right)=0$ with the co-moving time derivative (Lagrange form) where $\delta r$ is the distance between two neighbouring layers. It implies
\begin{equation}
\label{density}
\dot\rho=-2\rho\frac{\dot r}{r}-\rho\frac{\dot{\delta r}}{\delta r}
\end{equation}
for the density of a co-moving layer. Let us calculate the time derivatives at $t\to 0$. In Newtonian case we have ${\dot r}(t)=-\frac43 \pi G\rho(0) r(0)t+{\mathcal O}(t^3)$ and ${\dot{\delta r}}(t)=-\frac43 \pi G\rho(0) \delta r(0)t+{\mathcal O}(t^3)$ which gives
$${\dot\rho}(t)=4\pi G\rho^2(0)t+{\mathcal O}(t^3)$$
independent of $r$. The MOND law (\ref{MONDfall}) implies ${\dot r}(t)=-\sqrt{\frac43 \pi G\rho(0) a_0 r(0)}\ t+{\mathcal O}(t^3)$ and therefore ${\delta\dot r}(t)=-\sqrt{\frac{1}{3r(0)} \pi G\rho(0) a_0 }\ \delta r(0)t+{\mathcal O}(t^3)$. We see that according to
$$\dot\rho(t)=\sqrt{\frac{25\pi G\rho^3(0) a_0}{3r(0)}}\ t+{\mathcal O}(t^3)$$
the density grows faster in the central parts. Dynamics is different from what is expected in Newtonian physics. The central parts can become extremely dense well before one would say that the structure has formed as a whole for the full (uniform) density fluctuation of a large initial radius.

The outer layers of a more realistically distributed fluctuation also behave (expectedly) different. In this case, we assume that the mass of the outer layers is negligible, and only the full mass $M$ is relevant. In this case the Newtonian physics gives $\dot r(t)=-\frac{GM}{r^2(0)}t+{\mathcal O}(t^3)$ and $\dot{\delta  r}(t)=\frac{2GM}{r^3(0)}\delta r(0)t+{\mathcal O}(t^3)$. Substituting this into (\ref{density}) we get $\dot\rho={\mathcal O}(t^3)$. The density of outer layers starts growing only as a third order in time effect (if the initial velocities vanish). In MOND we have $\dot r(t)=-\frac{\sqrt{GMa_0}}{r(0)}t+{\mathcal O}(t^3)$ and  $\dot{\delta r}(t)=\frac{\sqrt{GMa_0}}{r^2(0)}\delta r(0)t+{\mathcal O}(t^3)$. It exhibits the growth of density as
$$\dot\rho(t)=\frac{\sqrt{GMa_0}}{r^2(0)}\ \rho(0) t+{\mathcal O}(t^3).$$

So, in MOND the internal parts collapse faster than the intermediate layers even if the density was nearly constant throughout. At the same time, the outer layers become denser more efficiently than in Newtonian gravity. If the ball is large enough and dense in its central part, then it has a Newtonian region at intermediate values of the radius. It clearly distinguishes MOND since such an effect is impossible with Dark Matter. The newtonised layers do not feel any additional mass inside the core of the cloud as if it was screened by the same amount of {\it negative} mass between the core and the given layer. This is nothing but a particular case  of the negative phantom mass prediction \cite{negativeM}. In spherical symmetry, we see that it is the $\sqrt{M}$ dependence of the force which is responsible for that. Then, during the collapse the inner MONDian layers would also enter the Newtonian regime, so that finally the central part would practically become Newtonian with MONDian halo around. MOND vastly increases efficiency of the gravitational collapse for fluctuations with very low Newtonian acceleration.

\subsection{Living with galaxies inside}

Of course, we can construct a spherical cloud of very complicated stratified profile. Then it would be possible to have a whole pile of Newtonian and MONDian layers in it. It is enough to 
surround the MONDian halo with a new massive spherical layer driving the force to Newtonian regime, and then to do the same around the new MONDian halo, et cetera. Note that averaging over large radial intervals might then yield the density profile dynamics wrong.

Admittedly, this is a rather academic problem. However, the real issue is even more complicated. The small scale structures (galaxies) generically form earlier than their large scale cousins. Suppose we have a large scale MONDian density fluctuation collapsing as a whole with smaller scale overdensities randomly distributed inside, well within their Newtonian regimes. So, the inner parts of the galaxies are Newtonian. Which attractive force should they experience towards the center of the proto-cluster? They apparently self-newtonise by pushing the argument of the $\mu(x)$ function to large values. It is hard to discuss such non-symmetric situations in canonical MOND framework of (\ref{modPoisson}), except the center of mass motion of a small body \cite{BekensteinMilgrom}. However, in simple MOND the procedure is obvious. We first solve for the Newtonian potential, and then find the actual acceleration according to relation (\ref{simpleMOND}). The Newtonian theory is linear, and $a_N$ is the sum of external force towards the centre of the cloud and internal attraction to the centre of the galaxy. The resulting $a_N$ is large in the core, and therefore the latter feels the Newtonian force. Meanwhile, the outer layers of the galaxy and the surrounding matter are in the deep MOND regime, and have to move towards the center of the cloud with larger acceleration. A tension appears between the inner and outer parts of galaxies with inner parts tending to fall slower. Probably, the galaxies are to be stripped off their outer layers, at least from the central side in the cloud, and the peripheral sides may start accumulating the faster matter from the deep MOND region.

In order to check the consequences in numerical simulations, an extremely good resolution is needed to cover several scales simultaneously. In a typical simulation \cite{Angus}, cubic cellular grids are used, with several hundred lattice points along each axis at distances of several Mpc from each other, the particles have galaxy-scale masses, and the gravitational forces are computed in finite differences of the potential values taken half a cell away from the points. Therefore, the galaxies are treated as point masses with no self-newtonisation effect. This is, of course, not the full story, or even totally incorrrect in simple MOND. 

And the same also concerns the structure of virialised galaxy clusters. The individual galaxies in the MONDian parts of a cluster should be in a complicated interaction with the external fields. It is unclear how strongly it can influence the structure of the cluster. This issue requires further investigation, and for now we leave it as an open problem.

\subsection{The problem of averaging and self-interaction}

We see that spatial averaging may change the dynamics in MOND because the interior parts of a massive object are screened from the MONDian effects by its own gravitational field. And this self-newtonising behaviour is lost upon averaging. In canonical MOND (\ref{modPoisson}) the conservation laws allow to prove \cite{BekensteinMilgrom} that, in the first approximation, the center of mass motion of a body with a small size and a small mass is the same as it would be for a point-like test particle. In simple MOND (\ref{simpleMOND}) it is not so, and the situation is even more delicate because an extended object will generically produce a non-zero self-force in an inhomogeneous external field since the values of $x$ in $\mu(x)$ would differ on opposite sides of the object.

An uncomfortable question arises as to which smallest scale do we have to go tracing the self-newtonising hierarchy. A point-like particle always produces an infinite field at its own position, and therefore would always feel the Newtonian gravity. If we consider general-relativistic Black Holes instead, then, as it can directly be seen from the fact that $a_0$ is approximately of order $cH_0$, the surface gravity at the horizon will turn as small as $a_0$ only for masses of order the whole mass in the visible part of the Universe. Therefore, the central Black Holes in galaxies are in this sense Newtonian. 

As another idea, we can use the fact that a particle can not be localised with a precision higher than its Compton wavelength $r_C\equiv\frac{h}{mc}$. Solving for a deep MOND regime $\frac{Gm}{r^2_C}<a_0$, we have\footnote{Under different reasonings, this mass value $m_{\star}$ has been mentioned in many papers on the subject including the very first one \cite{MOND}.}
$$m<m_{\star}=\sqrt[3]{\frac{a_0 h^2}{G c^2}}\approx\sqrt[3]{\frac{1.2\cdot 10^{-8}\ \rm\frac{cm}{s^2}\cdot\left(6.63\cdot 10^{-27}\ {\rm\frac{g\cdot cm^2}{s^2}}\right)^2}{6.67\cdot 10^{-8}\ {\rm\frac{cm^3}{s^2\cdot g}}\cdot\left(3\cdot 10^{10}\ {\rm\frac{cm}{s^2}}\right)^2}}\approx 2\cdot 10^{-25}\ \rm g.$$
Comparing $m_{\star}$ to the electron mass $m_e\approx 9.11\cdot 10^{-28}\ \rm g$ and the proton mass $m_p\approx 1.67\cdot 10^{-24}\ \rm g$, we see that an electron never breaks the MOND regime by its own self-field, while a proton can self-newtonise if localised sharply. Therefore, those sharply localised protons would behave differently as compared to the surrounding electrons, and must produce non-compensated electric currents. One might even speculate about contributing to the problem of primordial magnetic fields via those currents.

\section{Unavoidable non-linearity}

The MONDian regime is essentially non-linear. We have seen that, in spherical symmetry, a spherical shell does not produce an acceleration field inside, precisely as in General Relativity. Introducing a test particle of a very small but finite mass slightly breaks this statement in both models because the mass disturbs the shell and suffers a small acceleration. It turns out that in General Relativity it is much stronger suppressed than in MOND \cite{Dai} which is not surprising. Note that the Newtonian force of a spherical shell is compensated inside the shell because the force law $\sim\frac{1}{r^2}$ nicely fits the area growth $\sim r^2$ inside a given solid angle. With MOND, the force goes as $\frac{1}{r}$, and the non-linearity must precisely compensate for the unbalanced character of the force law and geometry. It works due to the Stokes' theorem for the equation (\ref{modPoisson}).

On the other hand, this non-linear behaviour in the {\it weak} field limit might seem problematic and unusual. It is not to say it's impossible. After all, it is realised in {\it TeVeS} with a condensate of additional degrees of freedom. The vacuum becomes a complicated physical medium, a kind of new incarnation of an aether. One can anticipate some problems such as Lorentz-violating effects around the condensates. And, what is even more important, despite all the great successes we do not (yet) have a fully satisfactory cosmological model with MOND. It calls for possible revision of the non-relativistic realisations, too. We see that the modified Poisson equation (\ref{modPoisson}) is very peculiar among all conceivable constructions in that it enjoys the Stokes' theorem. The price to pay is non-linearity. But how disastrous would it be to have a linear $\frac{1}{r}$  interaction in the large distance limit?

We can not just require the superposition principle for the
$$a=\frac{\sqrt{GMa_0}}{r}$$
acceleration fields with the $\sqrt{M}$ dependence. If we divide a gravitating body into small parts with such a law for each of the peaces, and apply the superposition principle, and go to the limit of infinitely fine-grained division, then the gravitational force would simply diverge. A straightforward way out is to introduce a new dimensionful constant, a characteristic mass $M_g$ of galactic size, and postulate for a point mass $M$ that $|\nabla\phi|=M\left(\frac{G}{r^2}+\sqrt{\frac{Ga_0}{M_g}}\cdot\frac{1}{r}\right)$. This is, of course, unsatisfactory because it contradicts the universal nature of the $a_0$ constant in galaxy dynamics. However, we can imagine that at galactic scales some non-linear physics is in operation which produces the proper MONDian laws, and then, at larger scales, it asymptotes to {\it linear} $\frac{1}{r}$ interactions. It might even ameliorate the situation with galaxy clusters in pure MOND.

Consider now a spherical shell of density $\rho$ and width $\Delta R$ at radius $R$ from the centre, and a test particle at radius $r$. The force in a small solid angle $\alpha$ in direction opposite to the centre would be
$$\sqrt{\frac{Ga_0}{M_g}}\cdot \rho\alpha\Delta R\cdot  (R-r),$$
and in direction to the centre
$$\sqrt{\frac{Ga_0}{M_g}}\cdot \rho\alpha\Delta R\cdot  (R+r).$$
In a very rough approximation, we conclude that the particle would be pulled towards the centre with an acceleration of order
$$a\sim 8\pi r\sqrt{\frac{Ga_0}{M_g}}\rho\Delta R,$$
independent of R. Naively, in an infinite Universe the resultant force diverges.

It seems reasonable to cut the integral over $R$ at the Hubble radius $R_H=\frac{c}{H_0}$. Then, for a galaxy of typical mass $M\approx M_g$ and typical radius $R\approx\sqrt{\frac{GM_g}{a_0}}$, we get the external force $a_{ext}\approx8\pi G\rho R_H$ which, with the baryonic density $\rho=\Omega_b\frac{3H^2_0}{8\pi G}$, gives $a_{ext}\approx 3\Omega_b H_0 c$. Taking $H_0\approx 0.07\ \rm\frac{km/s}{kpc}$, $c\approx 3\cdot 10^5 \ \rm km/s$, and $\Omega_b$ of a per cent order, we get the acceleration around $\sim 10^2\ \rm\frac{(km/s)^2}{kpc}$. Recall that $a_0$ is very close to that. Rotational velocities of a few hundreds kilometers per second at radii of a few tens of kiloparsecs give $a_0\sim 10^3\ \rm\frac{(km/s)^2}{kpc}$. The external force produces a considerable effect on the internal affairs of a galaxy. One might wish to deduce a sort of the Mach's principle out of that. In our opinion, it looks problematic and would probably have led to appearance of crystallographic type order in the large scale structure of the Universe. If so, it means that the modified Poisson equation (\ref{modPoisson}) is virtually the only viable non-relativistic realisation of MOND.

\section{Conclusions}

Recently, we have witnessed very interesting attempts to upgrade the phenomenological laws of MOND to a full cosmological model. This is a very complicated task which has to be attacked from different sides. And we are definitely in need of a precise and unequivocal relativistic formulation of MOND. 

Meanwhile, some conclusions can be, and must be, derived from the basic tenets of MOND. As we have discussed, there is no much freedom in writing down a non-relativistic model of MOND. Therefore it is worth studying it, even in absence of a full-fledged theory. We see that the properties of gravitational collapse are very unusual in MOND. And the non-linear character of equations might produce some (undesirable) effects when a whole hierarchy of scales is considered simultaneously. 

Moreover, there is a very pressing question about the smallest scale to which we need to resolve the dynamics. It would be very sad if we really need to discuss the self-newtonising effects all the way down to elementary particles. And, obviously, the locus of $|\nabla\phi|=0$ is very special for the differential equation (\ref{modPoisson}). Is the Cauchy problem for particle motion correctly defined with any practical level of predictability? We conclude that further analytic and numerical studies of MONDian dynamics are necessary.

{\bf Acknowledgments.} AG  is supported by Russian Foundation for Basic Research Grant No. 12-02-31214, and by the Saint Petersburg State University grant No. 11.38.660.2013.

\end{document}